# Improved EEG Event Classification Using Differential Energy

A. Harati, M. Golmohammadi, S. Lopez, I. Obeid and J. Picone

Neural Engineering Data Consortium, Temple University
Philadelphia, Pennsylvania, USA
{amir.harati, meysam, silvia.lopez, obeid, picone}@temple.edu

*Abstract*— **Feature extraction for automatic classification of EEG signals typically relies on time frequency representations of the signal. Techniques such as cepstral-based filter banks or wavelets are popular analysis techniques in many signal processing applications including EEG classification. In this paper, we present a comparison of a variety of approaches to estimating and postprocessing features. To further aid in discrimination of periodic signals from aperiodic signals, we add a differential energy term. We evaluate our approaches on the TUH EEG Corpus, which is the largest publicly available EEG corpus and an exceedingly challenging task due to the clinical nature of the data. We demonstrate that a variant of a standard filter bank-based approach, coupled with first and second derivatives, provides a substantial reduction in the overall error rate. The combination of differential energy and derivatives produces a *24%* absolute reduction in the error rate and improves our ability to discriminate between signal events and background noise. This relatively simple approach proves to be comparable to other popular feature extraction approaches such as wavelets, but is much more computationally efficient.**

## I. INTRODUCTION

Electroencephalograms (EEGs) are used in a wide range of clinical settings to record electrical activity along the scalp. EEGs are the primary means by which neurologists diagnose brain-related illnesses such as epilepsy and seizures [1]. We have developed a system, known as AutoEEG[TM], that automatically interprets EEGs, and delivers high performance on clinical data [2]. An overview of the system is shown in Figure 1. It incorporates a traditional hidden Markov model (HMM) based system and uses two stages of postprocessing to produce epoch labels. An *N*-channel EEG is transformed into *N* independent feature streams using a standard sliding window based approach. These features are then transformed into EEG signal event hypotheses using a standard HMM recognition system [3]. These hypotheses are postprocessed by examining temporal and spatial context to produce epoch labels.

Epochs are typically *1* sec in duration, while features are computed every *0.1* secs using *0.2* sec analysis windows. These parameters were optimized experimentally [2] in a previous study. Neurologists review EEGs in *10* sec windows, and it is common that pattern recognition systems classify *1* sec epochs. We further divide these *1* sec epochs into *10* frames of *0.1* secs each so that we can model an epoch with an HMM.

The system detects three events of clinical interest [4]: (1) spike and/or sharp waves (SPSW), (2) periodic lateralized epileptiform discharges (PLED), and (3) generalized periodic epileptiform discharges (GPED). SPSW events are epileptiform transients that are typically observed in patients with epilepsy. PLED events are indicative of EEG abnormalities and often manifest themselves with repetitive spike or sharp wave discharges that can be focal or lateralized over one hemisphere. These signals display quasi-periodic behavior. GPED events are similar to PLEDs, and manifest themselves as periodic short-interval diffuse discharges, periodic long-interval diffuse discharges and suppression-burst patterns according to the interval between the discharges. Triphasic waves, which manifest themselves as diffuse and bilaterally synchronous spikes with bifrontal predominance, typically at a rate of *1-2* Hz, are also included in this class.

The system also detects three events used to model background noise: (1) artifacts (ARTF) are recorded electrical activity that is not of cerebral origin, such as those due to the equipment, patient behavior or the environment; (2) eye movement (EYEM) are common events that can often be confused with a spike; (3) background (BCKG) is used for all other signals.

These six classes were arrived at through several iterations of a study conducted with Temple University Hospital neurologists. Automatic labeling of these events allows a neurologist to rapidly search long-term EEG recordings for anomalous behavior. Performance requirements for this application are extremely aggressive. For the system to be clinically useful, detection rates for the three signal classes must be at least *95%* with a false alarm rate below *5%*. This is a challenge for clinical data because the recordings contain many artifacts that can easily be interpreted as spikes. Therefore, neurologists still rely on manual review of data in clinical applications.

Hence, a unique aspect of the work reported here is that we have used the TUH EEG Corpus [2] for evaluation. TUH EEG is the world's largest publicly available database of clinical EEG data, comprising more than *28,000* EEG records and over *15,000* patients. It represents the collective output from Temple

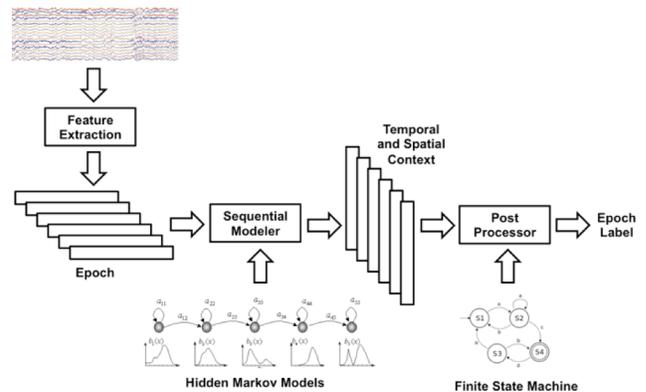

Figure 1. A two-level architecture for automatic interpretation of EEGs that integrates hidden Markov models for sequential decoding of EEG events with deep learning for decision-making based on temporal and spatial context.

University Hospital's Department of Neurology since 2002 and is an ongoing data collection project. EEG signals were recorded using several generations of Natus Medical Incorporated's Nicolet[TM] EEG recording technology. The raw signals obtained from the studies consist of multichannel recordings that vary between *20* and *128* channels sampled at a minimum of *250* Hz minimum using a *16*-bit A/D converter. The data is stored in a proprietary format that has been exported to EDF with the use of NicVue v5.71.4.2530. In our study, we have resampled all the data to a common sample frequency of *250* Hz.

## II. EEG FEATURES

Our system uses a fairly standard cepstral coefficient-based feature extraction approach similar to the Mel Frequency Cepstral Coefficients (MFCCs) used in speech recognition [3],[5],[6]. Though popular alternatives to MFCCs in EEG processing include wavelets, which are used by many commercial systems, our experiments with such features have shown very little advantage over MFCCs [7] on the TUH EEG Corpus. Therefore, in this study we have focused on filter bank approaches. Further, unlike speech recognition which uses a mel scale for reasons related to speech perception, we use a linear frequency scale for EEGs, since there is no physiological evidence that a log scale is meaningful [4].

The focus of this paper is an exploration of some traditional tuning parameters associated with cepstral coefficient approaches. In this study, we limit our explorations to the tradeoffs in computing energy and differential features, since these have the greatest impact on performance.

It is common in the MFCC approach to compute cepstral coefficients by computing a high resolution fast Fourier Transform, downsampling this representation using an oversampling approach based on a set of overlapping bandpass filters, and transforming the output into the cepstral domain using a discrete cosine transform [8],[9]. The zeroth-order cepstral term is typically discarded and replaced with an energy term as described below.

There are two types of energy terms that are often used: time domain and frequency domain. Time domain energy is a straightforward computation using the log of the sum of the squares of the windowed signal:

$$E_t = \log\left(\frac{1}{N}\sum_{n=0}^{N-1}|x(n)|^2\right)$$ (**Error! No sequence specified.**)

We use an overlapping analysis window (a 50% overlap was used here) to ensure a smooth trajectory of this features.

The energy of the signal can also be computed in the frequency domain by computing the sum of squares of the oversampled filter bank outputs after they are downsampled:

$$E_f = log(\sum_{k=0}^{N-1}|X(k)|^2)$$ (1)

This form of energy is commonly used in speech recognition systems because it provides a smoother, more stable estimate of the energy that leverages the cepstral representation of the signal. However, the virtue of this approach has not been extensively studied for EEG processing.

In order to improve differentiation between transient pulse-like events (e.g., SPSW events) and stationary background noise, we have introduced a differential energy term that attempts to model the long-term change in energy. This term examines energy over a range of *M* frames centered about the current frame, and computes the difference between the maximum and minimum over this interval:

$$E_d = \max_m\left(E_f(m)\right) - \min_m\left(E_f(m)\right)$$ (2)

We typically use a *0.9* sec window for this calculation. This simple feature has proven to be surprisingly effective.

The final step to note in our feature extraction process is the familiar method for computing derivatives of features using a regression approach [5],[8],[9]:

$$d_t = \frac{\sum_{n=1}^{N} n(c_{t+n}-c_{t-n})}{2\sum_{n=1}^{N} n^2}$$ (3)

where $d_t$ is a delta coefficient, from frame $t$ computed in terms of the static coefficients $c_{t+n}$ to $c_{t-n}$. A typical value for *N* is *9* (corresponding to *0.9* secs) for the first derivative in EEG processing, and *3* for the second derivative. These features, which are often called deltas because they measure the change in the features over times, are one of the most well-known features in speech recognition [8]. We typically use this approach to compute the derivatives of the features and then apply this approach again to those derivatives to obtain an estimate of the second derivatives of the features, generating what are often called delta-deltas. This triples the size of the feature vector (adding deltas and delta-deltas), but is well-known to deliver improved performance. This approach has not been extensively evaluated in EEG processing.

Dimensionality is something we must always pay attention to in classification systems since our ability to model features is directly related to the amount of training data available. The use of differential features raises the dimension of a typical feature vector from *9* (e.g., *7* cepstral coefficients, frequency domain energy and differential energy) to *27*. There must be sufficient training data to support this increase in dimensionality or any improvements in the feature extraction process will be masked by poor estimates of the model parameters (e.g., Gaussian means and covariances). As we will show in the next section, the TUH EEG Corpus is large enough to support such studies.

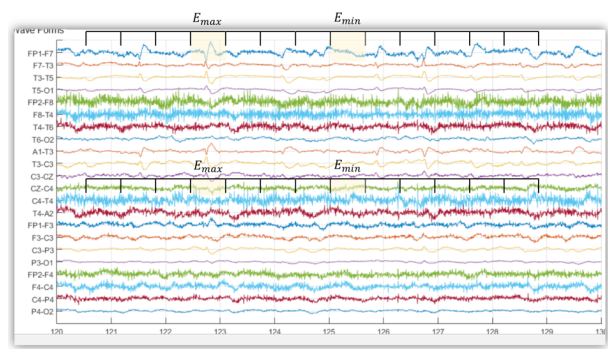

Figure 2. An illustration of how the differential energy term accentuates the differences between spike-like behavior and noise-like behavior. Detection of SPSW events is critical to the success of the overall system.

## III. EXPERIMENTATION

We have used a subset of TUH EEG that has been manually labeled for the six types of events described in Section I. The training set contains segments from *359* sessions while the evaluation set was drawn from *159* sessions. No patient appears more than once in the entire subset, which we refer to as the TUH EEG Short Set. A distribution of the frequency of occurrence of the *6* types of events in the training and evaluation set is shown in Table 1. The training set was designed to provide a sufficient number of examples to train statistical models such as HMMs. Note that some classes, such as SPSW, occur much less frequently in the actual corpus than common events such as BCKG. In fact, *99%* of the data is assigned to the class BCKG, so special care must be taken to build robust classifiers for the non-background classes. High performance detection of EEG events requires dealing with infrequently occurring events since the majority of the data is normal (uninformative). Hence, the evaluation set was designed to contain a reasonable representation of all classes.

We refer to the *6* classes shown in Table 1 as the *6*-way classification problem. This is not necessarily the most informative performance metric. It makes more sense to collapse the *3* background classes into one category. We refer to this second evaluation paradigm as a *4*-way classification task: SPSW, GPED, PLED and BACKG. The latter class contains an enumeration of the *3* background classes. Finally, in order that we can produce a DET curve [10], we also report a *2*-way classification task in which we collapse the data into a target class (TARG) and a background class (BCKG).

DET curves are generated by varying a threshold typically applied to likelihoods to evaluate the tradeoff between detection rates and false alarms. However, it is also instructive to look at specific numbers in table form. Therefore, all experiments reported in the tables use a scoring penalty of *0*, which essentially means we are evaluating the raw likelihoods returned from the classification system. In virtually all cases, the trends shown in these tables hold up for the full range of the DET curve.

### A. Absolute Features

The first series of experiments was run on a simple combination of features. A summary of these experiments is shown in Table 2. Cepstral-only features were compared with several energy estimation algorithms. It is clear that the combination of frequency domain energy and differential energy provides a substantial reduction in performance. However, note that differential energy by itself (system no. 4) produces a noticeable degradation in performance. Frequency domain energy clearly provides information that complements differential energy.

|  | Train | | Eval | |
|---|---|---|---|---|
| Event | No. | % (CDF) | No. | % (CDF) |
| SPSW | 645 | 0.8% ( 1%) | 567 | 1.9% ( 2%) |
| GPED | 6184 | 7.4% ( 8%) | 1,998 | 6.8% ( 9%) |
| PLED | 11,254 | 13.4% ( 22%) | 4,677 | 15.9% ( 25%) |
| EYEM | 1,170 | 1.4% ( 23%) | 329 | 1.1% ( 26%) |
| ARTF | 11,053 | 13.2% ( 36%) | 2,204 | 7.5% ( 33%) |
| BCKG | 53,726 | 63.9% (100%) | 19,646 | 66.8% (100%) |
| Total: | 84,032 | 100.0% (100%) | 29,421 | 100.0% (100%) |

Table 1. An overview of the distribution of events in the subset of the TUH EEG Corpus used in our experiments.

| No. | System Description | Dims. | 6-Way | 4-Way | 2-Way |
|---|---|---|---|---|---|
| 1 | Cepstral | 7 | 59.3% | 33.6% | 24.6% |
| 2 | Cepstral + $E_f$ | 8 | 45.9% | 33.0% | 24.0% |
| 3 | Cepstral + $E_t$ | 8 | 44.9% | 33.7% | 24.8% |
| 4 | Cepstral + $E_d$ | 8 | 55.2% | 32.8% | 24.3% |
| 5 | Cepstral + $E_f$ + $E_d$ | 9 | 39.2% | 30.0% | 20.4% |

Table 2. Performance on the TUH EEG Short Set of the base cepstral features augmented with an energy feature. System no. 5 uses both frequency domain and differential energy features. Note that the results are consistent across all classification schemes.

The improvements produced by system no. 5 hold for all three classification tasks. Though this approach increases the dimensionality of the feature vector by one element, the value of that additional element is significant and not replicated by simply adding other types of signal features [11].

### B. Differential Features

A second set of experiments were run to evaluate the benefit of using differential features. These experiments are summarized in Table 3. The addition of the first derivative adds about *7%* absolute in performance (e.g., system no. 6 vs. system no. 1). However, when differential energy is introduced, the improvement in performance drops to only *4%* absolute.

The story is somewhat mixed for the use of second derivatives. On the base cepstral feature vector, second derivatives reduce the error rate on the *6*-way task by *4%* absolute (systems no. 1, 6 and 11). However, the improvement for a system using differential energy is much less pronounced (systems no. 5, 10 and 15). In fact, it appears that differential energy and derivatives do something very similar. Therefore, we evaluated a system that eliminates the second derivative for differential energy. This system is labeled no. 16 in Table 3. We obtained a small but significant improvement in performance over system no. 10. The improvement on 4-way classification was larger, which indicates more of an impact on differentiating between PLEDs, GPEDs and SPSW vs. background. This is satisfying since this this feature was designed to address this problem.

| No. | System Description | Dims. | 6-Way | 4-Way | 2-Way |
|---|---|---|---|---|---|
| 6 | Cepstral + Δ | 14 | 56.6% | 32.6% | 23.8% |
| 7 | Cepstral + $E_f$ + Δ | 16 | 43.7% | 30.1% | 21.2% |
| 8 | Cepstral + $E_t$ + Δ | 16 | 42.8% | 31.6% | 22.4% |
| 9 | Cepstral + $E_d$ + Δ | 16 | 51.6% | 30.4% | 22.0% |
| 10 | Cepstral + $E_f$ + $E_d$ + Δ | 18 | 35.4% | 25.8% | 16.8% |
| 11 | Cepstral + Δ + ΔΔ | 21 | 53.1% | 30.4% | 21.8% |
| 12 | Cepstral + $E_f$ + Δ + ΔΔ | 24 | 39.6% | 27.4% | 19.2% |
| 13 | Cepstral + $E_t$ + Δ + ΔΔ | 24 | 39.8% | 29.6% | 21.1% |
| 14 | Cepstral + $E_d$ + Δ + ΔΔ | 24 | 52.5% | 30.1% | 22.6% |
| 15 | Cepstral + $E_f$ + $E_d$ + Δ + ΔΔ | 27 | 35.5% | 25.9% | 17.2% |
| 16 | (15) but no ΔΔ for $E_d$ | 26 | 35.0% | 25.0% | 16.6% |

Table 3. The impact of differential features on performance is shown. For the overall best systems (nos. 10 and 15), second derivatives do not help significantly. Differential energy and derivatives appear to capture similar information.

The results shown in Tables 1-3 hold up under DET curve analysis as well. DET curves for systems nos. 1, 5, 10, and 15 are shown in Figure 3. We can see that the relative ranking of the systems is comparable over the range of the DET curves. First derivatives deliver a measurable improvement over absolute features (system no. 10 vs. no. 5). Second derivatives do not provide as significant an improvement (system no. 15 vs. no. 10). Differential energy provides a substantial improvement over the base cepstral features.

It should be noted that user requirements for this type of technology includes an extremely low false alarm rate. Neurologists have expressed a need for a false alarm rate on the order of no more than one or two per day per bed while maintaining a detection rate of 95%. In related work we are able to approach these levels of performance using postprocessing steps alluded to in Figure 1. At these levels of performance, the differences between systems becomes more significant, and the use of second derivatives can potentially be more significant.

## IV. SUMMARY

In this paper, we have essentially calibrated some important algorithms used in feature extraction for EEG processing. We have shown that traditional feature extraction methods used in other fields such as speech recognition are relevant to EEGs. The use of a novel differential energy feature improved performance for absolute features (system nos. 1-5), but that benefit diminishes as first and second order derivatives are included. We have shown there is benefit to using derivatives and there is a small advantage to using frequency domain energy.

In related research [7],[11] we are evaluating approaches based on wavelets and other time-frequency representations. Preliminary results seem to indicate there are no significant benefits to these representations. Hence, in this work we have focused on optimization of our standard approach. Future work will focus on new feature extraction methods based on principles of deep learning [12], discriminative training [13] and nonparametric Bayesian models [14].

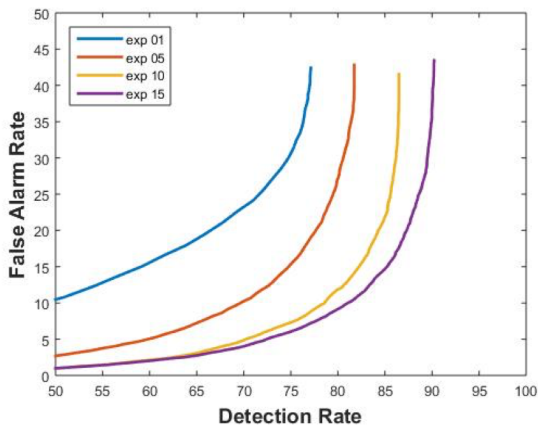

Figure 3. A DET curve analysis of feature extraction systems that compares absolute and differential features. The addition of first derivatives provides a measurable improvment in performance while second derivatives are less beneficial.

ACKNOWLEDGEMENTS

The primary funder of this research was the QED Proof of Concept program of the University City Science Center (Grant No. S1313). Research reported in this publication was also supported by the National Human Genome Research Institute of the National Institutes of Health under Award Number U01HG008468 and the National Science Foundation through Major Research Instrumentation Grant No. CNS-09-58854. The TUH EEG database work was funded by (1) the Defense Advanced Research Projects Agency (DARPA) MTO under the auspices of Dr. Doug Weber through the Contract No. D13AP00065, (2) Temple University's College of Engineering and (3) Temple University's Office of the Senior Vice-Provost for Research. Finally, we are also grateful to Dr. Mercedes Jacobson, Dr. Steven Tobochnik and David Jungries of the Temple University School of Medicine for their assistance in developing the classification paradigm used in this study and for preparing the manually annotated data.

REFERENCES

[1] T. Yamada and E. Meng, *Practical Guide for Clinical Neurophysiologic Testing: EEG*. Philadelphia, Pennsylvania, USA: Lippincott Williams & Wilkins, 2009.

[2] A. Harati, S. Lopez, I. Obeid, M. Jacobson, S. Tobochnik, and J. Picone, "THE TUH EEG CORPUS: A Big Data Resource for Automated EEG Interpretation," in *Proceedings of the IEEE SPMB*, 2014, pp. 1–5.

[3] J. Picone, "Continuous Speech Recognition Using Hidden Markov Models," *IEEE ASSP Mag.*, vol. 7, no. 3, pp. 26–41, Jul. 1990.

[4] S. Sanei and J. A. Chambers, *EEG signal processing*. Hoboken, New Jersey, USA: Wiley-Interscience, 2008.

[5] J. Lyons, "Mel Frequency Cepstral Coefficient (MFCC) tutorial," *Practical Cryptography*, 2015 (available: http://practicalcryptography.com/miscellaneous/machine-learning/guide-mel-frequency-cepstral-coefficients-mfccs/.

[6] S. Davis and P. Mermelstein, "Comparison of Parametric Representations for Monosyllabic Word Recognition in Continuously Spoken Sentences," *IEEE Trans. ASSP*, vol. 28, no. 4, pp. 357–366, 1980.

[7] P. Garrit, et al., "Wavelet Analysis for Feature Extraction on EEG Signals," presented at *Temple University CoE Res. Exp. for UG Conf.*, 2015 (available at http://www.isip.piconepress.com/publications/unpublished/conferences/2015/summer_of_code/wavelets/).

[8] X. Huang, A. Acero, and H.-W. Hon, *Spoken Language Processing: A Guide to Theory, Algorithm and System Development*. Upper Saddle River, New Jersey, USA: Prentice Hall, 2001.

[9] J. Picone, "Signal modeling techniques in speech recognition," *Proc. IEEE*, vol. 81, no. 9, pp. 1215–1247, 1993.

[10] A. Martin, G. Doddington, T. Kamm, M. Ordowski, and M. Przybocki, "The DET curve in assessment of detection task performance," in *Proceedings of Eurospeech*, 1997, pp. 1895–1898.

[11] A. Moura, I. Obeid, and J. Picone, "Feature Extraction Methods for EEG Event Detection," in *Temple University College of Eng. Res. Exp. for Undergrad. Conf.*, 2015 (available at http://www.isip.piconepress.com/publications/unpublished/conferences/2015/summer_of_code/features/).

[12] J. Snoek, R. P. Adams, and H. Larochelle, "Nonparametric guidance of autoencoder representations using label information," *J. Mach. Learn. Res.*, vol. 13, no. 1, pp. 2567–2588, 2012.

[13] D. Povey, et al., "Boosted MMI for model and feature-space discriminative training," *Proceedings of the IEEE Int. Conf. on ASSP,* Las Vegas, Nevada, USA, pp. 4057–4060, 2008.

[14] A. Harati and J. Picone, "A Doubly Hierarchical Dirichlet Process Hidden Markov Model with a Non-Ergodic Structure," submitted to the *IEEE/ACM Trans. Audio, Speech, Lang. Process.*, 2015.